\def\year{2017}\relax
\documentclass[letterpaper]{article} 
\usepackage{aaai17}  
\usepackage{times}  
\usepackage{helvet}  
\usepackage{courier}  
\usepackage{url}  
\usepackage{graphicx}  
\expandafter\def\expandafter\UrlBreaks\expandafter{\UrlBreaks
  \do\a\do\b\do\c\do\d\do\e\do\f\do\g\do\h\do\i\do\j%
  \do\k\do\l\do\m\do\n\do\o\do\p\do\q\do\r\do\s\do\t%
  \do\u\do\v\do\w\do\x\do\y\do\z\do\A\do\B\do\C\do\D%
  \do\E\do\F\do\G\do\H\do\I\do\J\do\K\do\L\do\M\do\N%
  \do\O\do\P\do\Q\do\R\do\S\do\T\do\U\do\V\do\W\do\X%
  \do\Y\do\Z}
\usepackage{floatrow}
\usepackage[inline]{enumitem}
\DeclareFloatFont{small}{\small}
\begin{document}
%
 \title{This Just In: Fake News Packs a Lot in Title, Uses Simpler, Repetitive Content in Text Body, More Similar to Satire than Real News}

\author{ 
Benjamin D. Horne and Sibel Adal{\i} \\
Rensselaer Polytechnic Institute\\
110 8th Street, Troy, New York, USA\\
\{horneb, adalis\}@rpi.edu
}

\maketitle
 \begin{abstract}
The problem of fake news has gained a lot of attention as it is
claimed to have had a significant impact on 2016 US Presidential Elections. Fake news
is not a new problem and its spread in social networks is
well-studied. Often an underlying assumption in fake news discussion
is that it is written to look like real news, fooling the reader who
does not check for reliability of the sources or the arguments in its
content. Through a unique study of three data sets and features that
capture the style and the language of articles, we show that this assumption is
not true. Fake news in most cases is more similar to satire than to
real news, leading us to conclude that persuasion in fake news is
achieved through heuristics rather than the strength of arguments. 
We show overall title structure and the use of proper nouns in
titles are very significant in differentiating fake from real. This
leads us to conclude that fake news is targeted for audiences who are
not likely to read beyond titles and is aimed at creating mental
associations between entities and claims.
 \end{abstract}

\section{Introduction}
The main problem we address in this paper is the following: Is there
any systematic stylistic and other content differences between fake
and real news?  News informs and influences almost all of our everyday
decisions. Today, the news is networked, abundant, and fast-flowing
through social networks. The sheer number of news stories together
with duplication across social contacts is overwhelming. The overload
caused by this abundance can force us to use quick heuristics to gain
information and make a decision on whether to trust its
veracity. These heuristics can come in many forms. In deciding that
content is believable by using reader's own judgment, readers may
simply skim an article to understand the main claims instead of
reading carefully the arguments and deciding whether the claim is
well-supported. In some cases, other heuristics may rely on the trust
for the source producing the information or for the social contact who
shared it. Often the trust decision is a combination of these
heuristics, content, source, and social network all playing a role. This trust decision strucure is supported by the well-studied notion of echo-chambers, in which the sharing  of information often conforms to one's beliefs and is impacted by homophily~\cite{bakshy2015exposure}. Furthermore, misleading or wrong
information have a higher potential to become viral
~\cite{Bessi:2015hg}and lead to negative
discussions~\cite{Zollo:2015hi}. Even more troubling, results show
resistance by individuals to information challenging established
beliefs and that attempts to debunk conspiracy theories are largely
ineffective~\cite{Anonymous:ZdbEGcdN}.

As an informed public is crucial to any operating democracy, incorrect
information is especially dangerous in political news.  As many widely
discredited claims have become viral and distributed widely in social
networks during the 2016 elections~\cite{buzzfeed}, the topic of
``fake news'' and effective methods for debunking it has gained
renewed attention. While there is a great deal of work on the social
factors leading to dissemination of misinformation in
networks~\cite{Bessi:2014ej},
there is relatively little work on understanding how fake news content
differs from real news content. We would like to understand whether
fake news differs systematically from real news in style and language
use. While some have argued that the distinction between fake and real
news can be a rather arbitrary one, as well-established news
organizations have been known to disseminate incorrect information on
occasion, such organizations operate with a degree of transparency
that is not found in fake news sites and risk their established
credibility if their news stories are shown to be false. More
importantly, their articles adhere to a journalistic style in making
and presenting claims.

To conduct our study of ``fake news'', we study three separate data
sets. Two of these data sets are novel: one has been featured by
Buzzfeed through their analysis of real and fake news items from 2016
US Elections~\cite{buzzfeed}. The second data set, collected by us,
contains news articles on US politics from real, fake, and satire news
sources. Finally, we look at a third data set containing real and
satire articles from a past study~\cite{burfoot2009automatic}. We
include satire as a type of fake news that relies on absurdity rather
than sound arguments to make claims, but explicitly identifies
itself as satire. Fake news in contrast has the intention to deceive,
making the reader believe it is correct. We study similarities between
fake news and satire to understand different heuristics that they both
employ to persuade their readers. The inclusion of satire as a third
category of news is a unique contribution of our work. Through a
statistical study of these three data sets, we show that fake news
articles tend to be shorter in terms of content, but use repetitive
language and fewer punctuation. Fake news articles differ much more in
their titles. Fake titles are longer, use few stop words, and fewer
nouns but more proper nouns. Furthermore, we find that fake news is
more similar to satire than real news. When fake news is different
than satire, the distinction simply exaggerates satire's differences
with real news further. Fake news packs the main claim of the article
into its title, which often is about a specific person and entity,
allowing the reader to skip reading the article, which tends to be
short, repetitive, and less informative. Given the arguments in the
article are less of a factor, the persuasion likely relies on
heuristics such as conformance of the information to one's beliefs. Lastly, we illustrate the predictive power of our features by utilizing linear kernel SVMs on small feature subsets. We hope that our study and data sets lead to further study of stylistic conventions used in persuading audiences with limited attention and
effective methods to counter them. In the least, it suggests that more
attention needs to paid to titles of such articles. 

\section{Related Work}
Fake news is certainly not a new phenomenon, and has been well studied
in both the fields of journalism and computer science. In particular,
it has been studied in two
ways: \begin{enumerate*}[label=(\arabic*)]\item analyzing the spread
of fake news and \item analyzing the content of fake
news.\end{enumerate*}

Rubin et al.~ work towards a news verification
system using Rhetorical Structure Theory (RST) on content data from
NPR's ``Bluff the Listener''~\cite{rubin2015towards}, achieving a 63\%
prediction accuracy over a 56\% baseline. Given its data source, this
study aims at identifying subtle differences in narrative between
different stories, which is not well-suited for news articles.
Similarly, Burfoot and Baldwin use SVMs on lexical and semantic
features to automatically classify true news content from satire news
content. The content features specifically geared towards satire
significantly outperform the baseline and achieve high classification
precision.~\cite{burfoot2009automatic}. Some of our features and one
of our data sets are common with this study, though we use a much
larger feature set that also captures the readability of the text and
its word usage. Along the same lines, Rubin et al.~ propose an SVM-based algorithm with 5 language features~\cite{rubin2016fake}. They achieve 90\% accuracy in detecting satire news from real news. Our features and classification method has some overlap with this work. However, these studies do not explicitly target
fake news. Yet, as demonstrated by the many events of the 2016 US
Election~\cite{buzzfeed}, fake and satirical news
are clearly different in motivation. Conversely, fake news is motivated
by deceiving its readers into thinking a completely false story is
real, many times with malicious intent~\cite{nprjestin}. This
distinction between the content of three classes of news: satire,
fake, and real, is the key contribution of our work.

The spread of misinformation in networks has also been
studied. Specifically, Bessi et al. study the attention given to
misinformation on Facebook. They show that users who often interact
with alternative media are more prone to interact with intentional
false claims~\cite{Bessi:2014ej}. Very
recently, Shao et al.~ launched a platform for tracking
online misinformation called
Hoaxy~\cite{Shao:2016:HPT:2872518.2890098}. Hoaxy gathers social news
shares and fact-checking through a mix of web scraping, web
syndication, and social network APIs. The goal of Hoaxy is to track
both truthful and not truthful online information
automatically. However, Hoaxy does not do any fact-checking of its
own, rather relying on the efforts of fact-checkers such
as \url{snopes.com}.

In our work, we concentrate on content analysis to study fake news for
several important reasons. Readers' assessment of information play a
significant role in decisions to disseminate it. Even though the
abundance of information leads to limited attention given to each
article, users engage with social media with the intention to find and
share information. When controlling for attention and individual
differences in the need to engage with information, the relevant
arguments and message characteristics become the determinant in
persuasion and attitude change~\cite{o2008elaboration}. Therefore, it
is helpful to understand whether there are specific the message
characteristics that accompany fake news articles being produced and
widely shared. As, textual analysis is well studied in computer science and
has been used for highly accurate document classification, it may
prove quite helpful in stopping the spread of fake
news~\cite{sebastiani2002machine}. Furthermore, we hope that this type
of analysis can be of benefit to grass-root fact-checkers by notifying
them of potential fake articles earlier. This benefit in turn can
provide more fact-checked content, ultimately helping systems like
Hoaxy.

\section{Methodology}
To begin exploring the content of fake news, we develop strict
definitions of what real, fake, and satire news stories
are. Specifically, real news stories are stories that are known to be
true and from well trusted news sources. Fake news stories are
stories that are known to be false and are from well known fake news
websites that are intentionally trying to spread
misinformation. Satire news stories are stories that are from
news sources that explicitly state they are satirical and do not
intentionally spread misinformation. Satire news is explicitly produced for
entertainment.
 
\subsection{Data sets}
With these definitions in mind, we use three independent data sets.

\subsubsection{Data set 1: Buzzfeed election data set}
First, we collected the news stories found in Buzzfeed's 2016 article
on fake election news on Facebook~\cite{buzzfeed}. Buzzfeed gathered
this data using the content analysis tool BuzzSumo by first searching
for real and fake stories getting the highest engagement on Facebook
using various methods during the 9 months before the 2016 US
Presidential Election, divided into three 3-month segments. For fake
stories, they targeted articles with the highest engagement for key
election terms and filtered these by known fake news sources. For real
stories, they found the stories getting the highest engagement from
well-known news organizations in the same time period. The URL and
Facebook engagement statistics of the chosen fake and real stories, 60
each, were made available. To keep the ground truth of real/fake
within our definitions, we filtered out stories that are opinion based
or were explicitly satire; leaving us with 36 real news stories and 35
fake news stories. Other than this filtering, we took the
ground truth as is; thus, it is important to keep in mind the
limitations of this data set. First, we do not know if there was any
selection bias in collecting the data, which could impact our
results. Second, while we can say these were news stories with
high user engagement, we cannot say anything about the actual traffic
the stories generated. Despite these limitations, this data set
provides reasonable ground truth labels and we know all stories  were
highly shared on social media.
 
\subsubsection{Data set 2: Our political news data set}
Given data set 1 contains political news only, we created our own
political news data set to strengthen our analysis and control for the
limitations of the first data set. Our data set contains 75 stories
from each of the three defined categories of news: real, fake, and
satire. We collected this data by first gathering known real, fake,
and satire new sources, which can be found in
Table~\ref{dataset2sources}. The fake news sources were collected from
Zimdars' list of fake and misleading news
websites~\cite{zimdar} and have had at least 1 story show up as false
on a fact-checking website like \url{snopes.com} in the past. The
real sources come from Business Insider's ``Most-Trusted''
list~\cite{bizins}, and are well established news media companies. The
satire sources are sites that explicitly state they are satirical on
the front page of their website. Once the sources were chosen, we
then randomly selected political stories from each of the sources. Each of these stories must be a ``hard'' news story, and not
an opinion piece.
The sources used in this data set have some overlap with data set 1,
but all collected stories are different than those in the first data
set. While we cannot say anything about the engagement of the articles
in data set 2, it avoids the possible limitations of the Buzzfeed data
set described above. Furthermore, data set 2 allows us to explicitly
analyze all three defined categories of news by having both satire and
fake news stories in the same data set.

Both data sets 1 and 2 will be publicly available together with this
paper at \url{https://github.com/rpitrust/fakenewsdata1}.

\subsubsection{Data set 3: Burfoot and Baldwin data set}
Finally, we use a data set from \cite{burfoot2009automatic}, which
consists of 233 satire news stories and 4000 real news stories used in
a classification task between these two types of news stories using
lexical and semantic features. The authors collect the real news
stories using newswire documents sampled from the English Gigaword
Corpus. To select satire documents, they hand select satire stories
that are closely related in topic to the real stories collected. They
manually filter out ``non-newsy'' satire, similar to the method we use
to construct our data set. While Burfoot and Baldwin do control for
topic when matching articles, they do not limit themselves to only
political topics; thus, this data may not be directly comparable to
our other two data sets with respect to some of our more topic-driven features. We include this data set to strengthen our
categorical comparisons between satire and real stories. A separate
limitation of Burfoot and Baldwin's data set is that we do not
explicitly know the sources of each story and cannot verify that the
definition of real news sources by the authors corresponds to ours.

\begin{table}[h]
\begin{tabular}{| c | c | c |}
\hline
\textbf{Real sources} & \textbf{Fake sources} & \textbf{Satire sources} \\
\hline
Wall Street Journal & Ending the Fed & The Onion \\
The Economist & True Pundit & Huff Post Satire \\
BBC & abcnews.com.co & Borowitz Report \\
NPR & DC Gazette & The Beaverton \\
ABC & libertywritersnews & SatireWire \\
CBS & Before its News & Faking News\\
USA Today & Infowars & \\
The Guardian & Real News Right Now &  \\
NBC & &  \\
Washington Post & &  \\
\hline
\end{tabular}
\caption{Data set 2 sources}
\label{dataset2sources}
\end{table}

\begin{table*}[t]
\centering
\begin{tabular}{ | c || c | c | c | c | }
\hline
\multicolumn{5}{|c|}{Metadata} \\
 \hline
 Data set ID & description & \# real news & \# fake news & \# satire news\\
 \hline
 1 & filtered Buzzfeed 2016 election data set & 36 & 35 & 0\\
 2 & our 3 class political news data set & 75 & 75 & 75\\
 3 & Burfoot and Baldwins satire data set & 4000 & 0 & 233\\
 \hline
 \end{tabular}
 \caption{\label{dataids} Ground truth counts and ID number of each data set used in this study.}
 \end{table*}

\begin{table*}[thb!]
\centering
\begin{minipage}[t]{3.2in}
\begin{tabular}{p{0.6in}p{2.2in}}  
  Abbr. & Description \\ \hline
  GI & Gunning Fog Grade Readability Index\\
  SMOG & SMOG Readability Index\\
  FK & Flesh-Kincaid Grade Readability Index\\
  med\_depth & median depth of syntax tree \\
  med\_np\_depth & median depth of noun phrase tree\\
  med\_vp\_depth & median depth of verb phrase tree\\
   flu\_coca\_c & avg. frequency of least common 3 words using all of the coca corpus \\
   flu\_coca\_d & avg. frequency of words in each document using all of the coca corpus \\
   TTR & Type-Token Ratio (lexical diversity) \\
   avg\_wlen & avg. length of each word \\
   \\ \hline \\
  \multicolumn{2}{c}{(a) Complexity Features} \\ \\ \\
	analytic & number of analytic words\\  	
  	insight & number of insightful words\\
  	cause & number of causal words\\
  	discrep & number of discrepancy words \\
  	tentat & number of tentative words \\
  	certain & number of certainty words \\
  	differ & number of differentiation words \\
  	affil & number of affiliation words \\
  	power & number of power words \\
  	reward & number of reward words \\
  	risk & number of risk words \\
  	personal & number of personal concern words (work, leisure, religion, money)\\
  	tone & number of emotional tone words\\
  	affect & number of emotion words (anger, sad, etc.)\\
  	str\_neg & strength of negative emotion using SentiStrength\\
  	str\_pos & strength of positive emotion using SentiStrength\\
    \\ \hline \\
     \multicolumn{2}{c}{(b) Psychology Features}  
\end{tabular}
  \end{minipage} 
\begin{minipage}[t]{3.2in}
\begin{tabular}{p{0.6in}p{2.4in}}  
  Abbr. & Description \\ \hline
  WC & word count\\
  WPS  & words per sentence\\
  NN  & number of nouns\\
  NNP & number of proper nouns\\
  PRP & number of personal pronouns\\
  PRP\$ & number of possessive pronouns\\
  WP & Wh-pronoun\\
  DT & number of determinants\\
  WDT & number of Wh-determinants\\
  CD & number of cardinal numbers\\
  RB & number of adverbs\\
  UH & number of interjections\\
  VB & number verbs\\
  JJ	& Adjective\\
  VBD & number of past tense verbs\\
  VBG &	Verb, gerund or present participle\\
 VBN	& Verb, past participle\\
 VBP	& Verb, non-3rd person singular present\\
 VBZ	& Verb, 3rd person singular present\\
  focuspast & number of past tense words\\
  focusfuture & number of future tense words\\
  i & number of I pronouns (similar to PRP)\\
  we & number of we pronouns (similar to PRP)\\
  you & number of you pronouns (similar to PRP)\\
  shehe & number of she or he pronouns (similar to PRP)\\
  quant & number of quantifying words\\
  compare & number of comparison words \\
  exclaim & number of exclamation marks\\
  negate & number of negations (no, never, not)\\
  swear & number of swear words \\
  netspeak & number of online slang terms (lol, brb)\\
  interrog & number of interrogatives (how, what, why)\\
  all\_caps & number of word that appear in all capital letters\\
  per\_stop & percent of stop words (the, is, on)\\
  allPunc & number of punctuation\\
  quotes & number of quotes\\
  \#vps & number of verb phrases\\
  \hline \\
  \multicolumn{2}{c}{(c) Stylistic Features} \\ 
\end{tabular}
\end{minipage}
\caption{\label{tbl:features} Different features used in our study}
\end{table*}

\subsection{Features}
To study these different articles, we compute many content based
features on each data set and categorize them into 3 broad categories:
stylistic, complexity, and psychological.

\subsubsection{Stylistic Features}
The stylistic features are based on natural language processing to
understand the syntax, text style, and grammatical elements of each
article content and title. To test for differences in syntax, we use
the Python Natural Language Toolkit~\cite{bird2006nltk} part of speech
(POS) tagger and keep a count of how many times each tag appears in
the article. Along with this, we keep track of the number of
stop-words, punctuation, quotes, negations (no, never, not),
informal/swear words, interrogatives (how, when, what, why), and words
that appear in all capital letters. For features that need word
dictionaries, such as the number of informal words, we use the 2015
Linguistic Inquiry and Word Count (LIWC)
dictionaries~\cite{tausczik2010psychological}.

\subsubsection{Complexity Features}
The complexity features are based on deeper natural language
processing computations to capture the overall intricacy of an article
or title. We look at two levels of intricacy: the sentence level and
the word level. To capture the sentence level complexity, we compute
the number of words per sentence and each sentence's syntax tree
depth, noun phrase syntax tree depth, and verb phrase syntax tree
depth using the Stanford Parser~\cite{stanparse}. We expect that more
words per sentence and deeper syntax trees mean the average sentence
structure complexity is high. 

To capture the word level complexity, we use several key
features. First, we compute the readability of each document using
three different grade level readability indexes: Gunning Fog,
SMOG Grade, and Flesh-Kincaid grade level index. Each measure
computes a grade level reading score based on the number of syllables
in words. A higher score means a document takes a higher education
level to read. Second, we compute what is called the Type-Token Ratio
(TTR) of a document as the number of unique words divided by
the total number of words in the document. TTR is meant to capture the
lexical diversity of the vocabulary in a document. A low TTR means a
document has more word redundancy and a high TTR means a document has
more word diversity~\cite{dillard2002persuasion}. Third, we compute a
word level metric called fluency, used in~\cite{horne2016expertise}.
Fluency is meant to capture how common or specialized the vocabulary
of a document is. We would say that a common term is more fluent, and
easier to interpret by others; while a less common term would be less
fluent and more technical. This idea is captured by computing how
frequently a term in a document is found in a large English corpus. We
use both the Corpus of Contemporary American English
(COCA)~\cite{coca} corpus to compute this feature.

\subsubsection{Psychological Features}
The psychological features are based on well studied word counts that
are correlated with different psychological processes, and basic
sentiment analysis. We use Linguistic Inquiry and Word Count (LIWC)
dictionaries~\cite{tausczik2010psychological} to measure cognitive
processes, drives, and personal concerns. Along with this, we use LIWC to measure basic bag-of-words sentiment. We then use SentiStrength~\cite{thelwall2010sentiment} to
measure the intensity of positive and negative emotion in each
document. SentiStrength is a sentiment analysis tool that reports a
negative sentiment integer score between -1 and -5 and a positive
sentiment integer score between 1 and 5, where -5 is the most negative
and 5 is the most positive.

\subsection{Statistical Tests and Classification}
Due to our use of small data sets and large number of
features, we choose to first approach the problem using two well known
hypothesis testing methods, the one-way ANOVA test and the Wilcoxon
rank sum test, to find which features differ between the different
categories of news.

A one-way ANOVA test is used to compare the means of different groups
on a dependent variable. ANOVA uses the ratio of the treatment and
residual variances to determine if the difference in group means is
due to random variation or the treatment. In our case, the treatment
is grouping news as fake, real, or satire. ANOVA assumes that the
variables are approximately normally distributed. While these
assumptions are true for most of our features, we cannot assume it is
true for all of them. Thus, we also utilize the Wilcoxon rank sum test
to compare two distributions that are not normal. Specifically, for each feature that
passes a normality test, we use the one-way ANOVA, otherwise, we will use Wilcoxon rank sum. In both cases,
we are looking for a large F-value and a significance level of at least 0.05.

These statistical test cannot say anything about predicting classes in the data,
as a machine learning classifier would. However, these test can
illustrate a shift in the use of a linguistic feature based on the
category the article falls in. We would expect that the more
significant the difference in a feature, the higher chance a
machine learning classifier would be able to separate the
data. To better test the predictive power of our features, we will use a linear classifier on a small subset of our features. We select the top 4 features from our hypothesis testing methods for both the body text and title text of the articles. With these 4 features, we will run a Support Vector Machine (SVM) model with a linear kernel and 5-fold cross-validation. Since we are only using a small number of features and our model is a simple linear model on balanced classes, we expect to avoid the over-fitting that comes with small data.

\begin{table}[t!]
\caption{Features that differ in the {\em body} of the news content (bolded results correspond to p values of 0.00 or less, all other results have p values of at least less than 0.05)}
\label{BodyResults}
\hspace*{-0.2in}\begin{tabular}{  c || c | c | c   }

 Feature & Data set 1 &  Data set 2 & Data set 3\\
 \hline
WC   & {\bf Real $>$ Fake} & {\bf Real $>$ Fake $>$ Satire} & Satire $>$ Real \\
flu\_coca\_c  & Fake $>$ Real & Satire $=$ Fake $>$ Real &    \\
flu\_coca\_d  &  & Satire $=$ Fake $>$ Real &    \\
flu\_acad\_c  &  & Satire $>$ Fake $=$ Real &    \\
avg\_wlen & Real $>$ Fake & {\bf Real $>$ Fake $=$ Satire} & Real $>$ Satire \\
quote & Real $>$ Fake & {\bf Real $>$ Fake $=$ Satire} & \\
allPunc & {\bf Real $>$ Fake} & Real $>$ Fake $=$ Satire  & \\
GI &  & Real $=$ Satire $>$  Fake & {\bf Satire $>$ Real} \\
FK &  & Real $=$ Satire $>$ Fake & {\bf Satire $>$ Real} \\
analytic &  & Real $>$ Fake $=$ Satire  & \\
all\_caps &  & Fake $=$ Satire $>$ Real  & \\
NN &  & Real $>$ Fake $=$ Satire & \\
PRP &  & {\bf Satire $>$ Fake $>$ Real} & Satire $>$ Real\\
PRP\$ &  & Satire $>$ Fake $=$ Real & Satire $>$ Real \\
DT &  & Fake $=$ Real $>$ Satire & \\
WDT &  & Fake $=$ Real $>$ Satire & \\
RB &  & {\bf Satire $=$ Fake $>$ Real} & \\
i &  & Satire $>$ Fake $=$ Real & Satire $>$ Real\\
we &  & Fake $>$ Real $=$ Satire & \\
you &  & Satire $>$ Fake $>$ Real & Satire $>$ Real\\
shehe	 &  & Satire $>$ Real $=$ Fake & Satire $>$ Real\\
CD	 &  & Real $=$ Fake $>$ Satire & {\bf Real $>$ Satire}\\
compare &  & Real $>$ Fake $=$ Satire & \\
swear	 &  & Satire $>$ Real $=$ Fake & \\
TTR & {\bf Fake $>$ Real} & {\bf Satire $>$ Fake $>$ Real} & {\bf Satire $>$ Real}\\
avg\_negstr & {\bf Fake $>$ Real} & & \\
avg\_posstr & Real $>$ Fake & & \\
med\_depth & Fake $>$ Real & & Satire $>$ Real\\
med\_np\_d & Fake $>$ Real & & \\
med\_vp\_d & Fake $>$ Real & & \\

\end{tabular}
\end{table}

\begin{table}[t!]
\caption{Features that differ in the {\em title} of the news content (bolded results correspond to p values of 0.00 or less, all other results have p values of at least less than 0.05)}
\label{TitleResults}
\hspace*{-0.2in}\begin{tabular}{ c || c | c | c   }

 Feature & Data set 1 &  Data set 2 & Data set 3\\
 \hline
WPS & Fake $>$ Real  & Fake $>$ Real $=$ Satire  & \\
flu\_coca\_c  & Fake $>$ Real  & {\bf Satire $>$ Fake $=$ Real} &  Satire $>$ Real  \\
avg\_wlen &  & {\bf Real $>$ Fake $=$ Satire} &  \\
all\_caps	& Fake $>$ Real & Satire $>$ Fake $>$ Real &  \\
GI	 & Real $>$ Fake & {\bf Real $>$ Satire $=$ Fake}  & Real $>$ Satire \\
FK & Real $>$ Fake & {\bf Real $>$ Satire $=$ Fake}  & {\bf Real $>$ Satire} \\
NNP & {\bf Fake $>$ Real} & {\bf Fake $=$ Satire $>$ Real}  & {\bf Satire $>$ Real} \\
NN	 & Real $>$ Fake & {\bf Real $>$ Satire $>$ Fake}  &  \\
PRP & Real $>$ Fake &   &  Satire $>$ Real\\
PRP\$ & Real $>$ Fake  &   & Satire $>$ Real \\
DT & Real $>$ Fake  &   & \\
CD &  &   & Real $>$ Satire \\
per\_stop & {\bf Real $>$ Fake}  & {\bf Real $>$ Satire $=$ Fake}  &  {\bf Real $>$ Satire}\\
exclaim &  & Fake $>$ Real $=$ Satire &  \\
focuspast &  & {\bf Fake $>$ Satire $=$ Real} &  \\
analytic & {\bf Fake $>$ Real} & &\\
\#vps & & {\bf Fake $>$ Real = Satire} & \\

\end{tabular}
\end{table}

\section{Results}
In this section, we present the most significant features in
separating news,  referring to each data set by its ID
number found in Table~\ref{dataids}. The complete results can be found
in Tables~\ref{BodyResults} and~\ref{TitleResults}. Classification
results can be found in Table~\ref{svm}.

\paragraph{The content of fake and real news articles is substantially different.}
Our results show there is a significant difference in the content of
real and fake news articles. Consistently
between data sets 1 and 2, we find that real news articles are
significantly longer than fake news articles and that fake news
articles use fewer technical words, smaller words, fewer punctuation,
fewer quotes, and more lexical redundancy. Along with this, in data set
2, fake articles need a slightly lower education level to read, use
fewer analytic words, have significantly more personal pronouns, and
use fewer nouns and more adverbs. These differences illustrate a strong
divergence in both the complexity and the style of content. Fake news
articles seem to be filled with less substantial information
demonstrated by having a high amount of redundancy, more adverbs, fewer
nouns, fewer analytic words, and fewer quotes. Our results also suggest that fake news may be more personal and more
self-referential, using words like we, you, and us more
often. However, this result is not consistent between data sets and is
less significant.

This stark difference between real and fake news content is further strengthened by our SVM classification results on the content of fake and real articles in Dataset 2. To classify the content, we use the top 4 features from our statistical analysis: number of nouns, lexical redundancy (TTR), word count, and number of quotes. We achieve a 71\% cross-validation accuracy over a 50\% baseline when separating the body texts of real and fake news articles. Similarly, when classifying fake from real content in Data set 1 using the same 4 features, we achieve a 77\% accuracy over a 57\% baseline. These results are shown in Table~\ref{svm}.

\begin{table}[h]
\centering
\hspace*{-0.15in}\begin{tabular}{ | c | c || c | c | c | }
\hline
  & Baseline & Fake vs Real & Satire vs Real & Satire vs Fake \\
 \hline
Body& 50\% &  71\% & 91\% & 67\%\\
Title& 50\% &  78\% & 75\% & 55\%\\
 \hline
 \end{tabular}
 \caption{\label{svm} Linear kernel SVM classification results using the top 4 features for the body and the title texts in Data set 2. Accuracy is the mean of 5-fold cross-validation.}
 \end{table}

  \vspace*{-0.2in}

\paragraph{Titles are a strong differentiating factor between fake and real news.}
When looking at just the titles of fake and real news articles, we
find an even stronger dissimilarity between the two, with high
consistency between the data sets and high statistical significance in
the differences. Precisely, we find that fake news titles are longer than
real news titles and contain simpler words in both length and
technicality. Fake titles also used more all capitalized words,
significantly more proper nouns, but fewer nouns overall, and fewer
stop-words. Interestingly, we also find that in data set 1 fake titles
use significantly more analytical words and in data set 2 fake titles
use significantly more verb phrases and significantly more past tense
words. Overall, these results suggest that the writers of fake news
are attempting to squeeze as much substance into the titles as
possible by skipping stop-words and nouns to increase proper nouns and
verb phrases.

Looking at an example from our data will solidify this notion:

\begin{enumerate}
\item \textbf{FAKE TITLE:} "BREAKING BOMBSHELL: NYPD Blows Whistle on New Hillary Emails: Money Laundering, Sex Crimes with Children, Child Exploitation, Pay to Play, Perjury"

\item \textbf{REAL TITLE:} Preexisting Conditions and Republican Plans to Replace Obamacare
\end{enumerate}

There is a clear difference in the number of claims being made in each
title. The fake title uses many verb phrases and name entities to get
many points across, while the real title opts for a brief and general
summary statement. This broad pattern shown in the example is
consistent across the first two data sets. Interestingly, this finding in the titles of fake news is closely related to Rubin et al's finding in the quotations of satirical articles~\cite{rubin2016fake} where a large number of clauses are packed into a sentence. Saliently, this title structure is not as similar to clickbait as one might expect. Clickbait titles have been shown to have many more function words, more stopwords, more hyperbolic words (extremely positive), more internet slangs, and more possestive nouns rather than proper nouns~\cite{chakraborty2016stop}.

Once again, these differences are further strengthened by our SVM classification results on Data set 2. To classify fake and real news articles by their title, we use the top 4 features from our statistical analysis on titles: the percent of stopwords, number of nouns, average word length, and FKE readability. We achieve a 78\% cross-validation accuracy over a 50\% baseline, demonstrating the importance of the title in predicting fake and real news. These results are shown in Table~\ref{svm}. In addition, when classifying fake from real titles in Data set 1 using the same 4 features, we achieve a 71\% accuracy over a 49\% baseline.

Table~\ref{titlefigtab} shows the distributions of select features
that are significant and consistent between data sets.
 
\begin{table}[ht]
  \centering
  \begin{tabular}{cc}
    Data set 1 & Data set 2 \\
    \includegraphics[width=4cm]{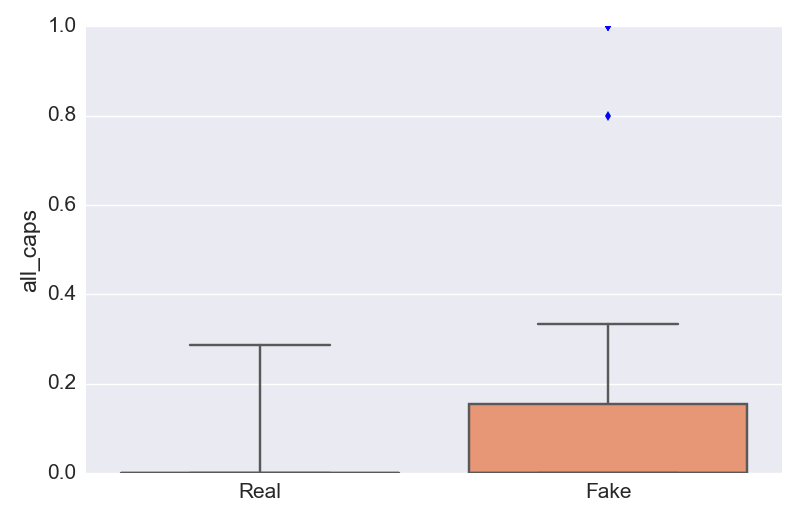} & \includegraphics[width=4cm]{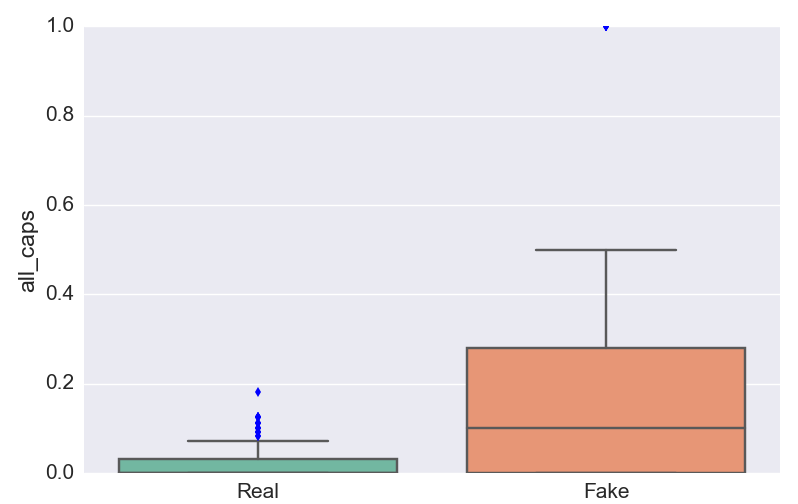} \\
    \includegraphics[width=4cm]{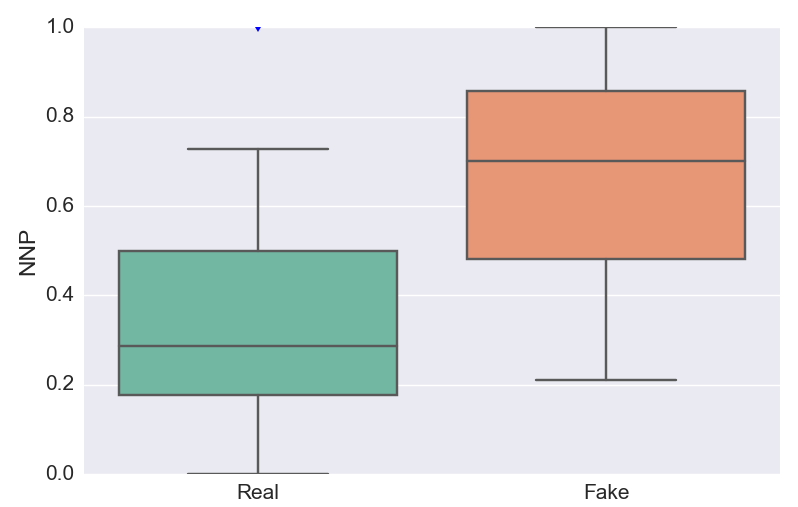} & \includegraphics[width=4cm]{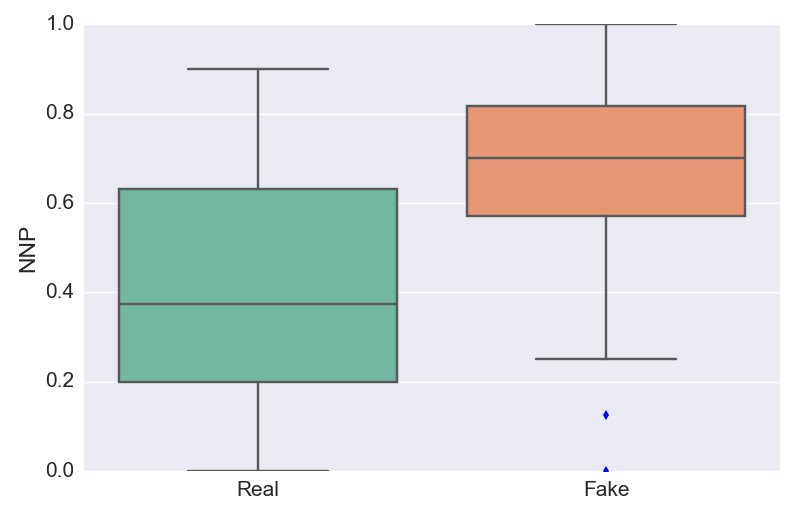} \\
    \includegraphics[width=4cm]{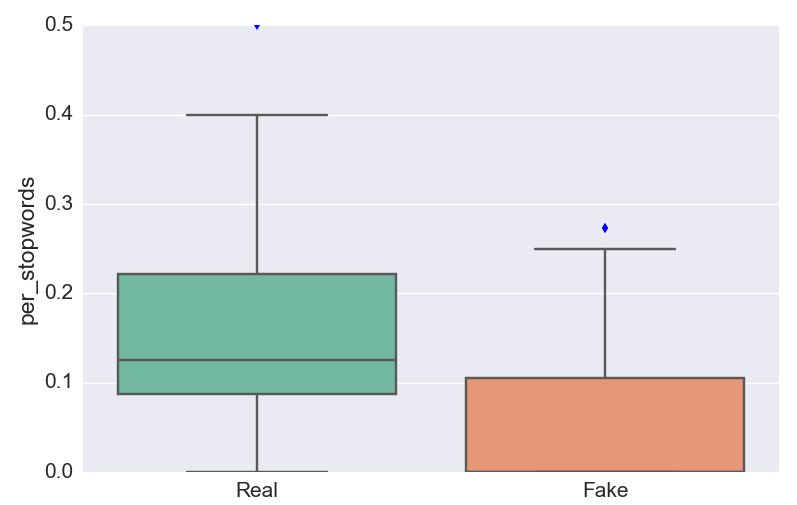} & \includegraphics[width=4cm]{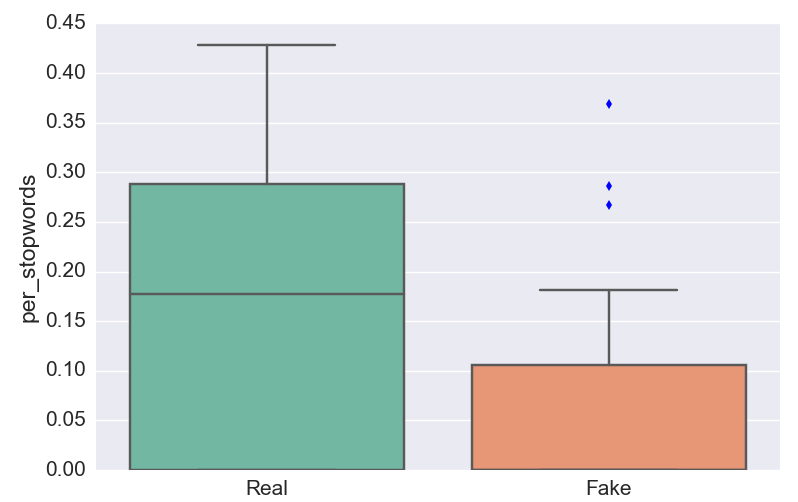} \\
  \end{tabular}
  \caption{95\% Confidence plots of all\_caps, NNP, and per\_stop. These features were found to be significant across both data sets 1 and 2 for fake and real titles. 
  \textbf{Top:} all\_caps \textbf{Middle:} NNP \textbf{Bottom:} per\_stop}\label{titlefigtab}
\end{table}

  \vspace*{-0.2in}

\paragraph{Fake content is more closely related to satire than to real.}
When adding in satire articles to the analysis, we find that the
majority of our features distributions are common between satire and
fake. Specifically, both satire and fake use smaller, fewer technical,
and fewer analytic words, as well as, fewer quotes, fewer punctuation,
more adverbs, and fewer nouns than real articles. Further, fake and
satire use significantly more lexical redundancy than real articles.

These claims are further supported by SVM classification results
in Table~\ref{svm}. When using the number of nouns, lexical redundancy
(TTR), word count, and number of quotes for the body text, we achieve
a 91\% cross-validation accuracy over a 50\% baseline on separating
satire  from real articles. On the other hand, when separating
satire  from fake articles, we only acheive a 67\% accuracy
over a 50\% baseline. Similarly, when classifying the titles of
articles using the percent of stopwords, number of nouns, average word
length, and FKE readability, we achieve a 75\% cross-validation
accuracy when separating satire titles from real titles, but only
achieve a 55\% accuracy when separating satire titles from fake
titles. While this accuracy is a reasonable improvement over baseline,
it is not nearly as high as big of an improvement as separating satire
from real or real from fake.

Overall, these results paint an interesting picture where satire and fake news
articles are written in a less investigative way. This conclusion is
in sync with what we know of
satire~\cite{randolph1942structural}~\cite{burfoot2009automatic}. Satire
news articles do not have the goal of creating sound arguments, but
often make absurd and eccentric claims, whereas real news articles
must back up the information they are providing with direct quotes,
domain specific knowledge, and reasonable analysis.

\paragraph{Real news persuades through arguments, while fake news persuades through heuristics.}
To better explain our findings, we look at the Elaboration Likelihood
Model (ELM) of persuasion, well-studied in communications. According
to ELM, people are persuaded through two different routes: 
central and peripheral~\cite{petty1986}. The central
route of persuasion results from the attentive examination of the
arguments and message characteristics presented, involving a
high amount of energy and cognition. In opposition, the peripheral
route of persuasion results from associating ideas or making
conjectures that are unrelated to the logic and quality of the
information presented. This method could  be called a heuristic
method as it does not ensure to be optimal or even sufficient in
achieving its objective of finding correct or truthful
information. The peripheral route takes very little energy and
cognition.

This model fits well with both our results and recent studies on the
behavior of sharing online information. Given the similarity of fake
content to satire, we hypothesize that fake content targets the
peripheral route, helping the reader use simple heuristics to assess
veracity of the information.  The significant features support this
finding: fake news places a high amount substance and claims into
their titles and places much less logic, technicality, and sound
arguments in the body text of the article. Several studies have argued
that the majority of the links shared or commented on in social
networks are never clicked, and thus, only the titles of the articles
are ever
read~\cite{wang2016measuring}
Titles of fake news often present claims about people and entities in
complete sentences, associating them with actions. Therefore, titles
serve as the main mechanism to quickly make claims which are easy to
assess whether they are believable based on the reader's existing
knowledge base. The body of fake news articles add relatively little
new information, but serves to repeat and enhance the claims made in
the title. The fake content is more negative in general, similar to
findings in past work~\cite{Zollo:2015hi}. Hence, fake news is assessed through
the peripheral route. We hypothesize that if users were 
to utilize the central route of persuasion, they would have a much
lower chance of being convinced by the body content of fake news, as
we have shown the fake news content tend to be short, repetitive and
lacking arguments. 

\section{Conclusions and Future work}
In this paper, we showed that fake and real news articles are notably
distinguishable, specifically in the title of the articles. Fake news
titles use significantly fewer stop-words and nouns, while using
significantly more proper nouns and verb phrases. We also conclude
that the complexity and style of content in fake news is more closely
related to the content of satire news. We also showed that our
features can be used to significantly improve the prediction of fake
and satire news, achieving between 71\% and 91\% accuracy in
separating from real news stories. These results lead us to consider
the Elaboration Likelihood Model as a theory to explain the spread and
persuasion of fake news. We conclude that real news articles persuade
users through sound arguments while fake news persuades users through
heuristics. This finding is concerning as a person may be convinced of
fake news simply out of having low energy, not just contempt,
negligence, or low cognition. Unfortunately, misleading claims in the
titles of fake news articles can lead to established beliefs which can
be hard to change through reasoned arguments. As a starting point,
articles that aim to counter fake claims should consider packing
the counter-claim into their titles.

This work has some limitations and room for improvement. First, we
would like to extensively expand the news data set. Typically it is
very difficult to obtain a non-noisy ground truth for fake and real
news, as the real news is becoming increasingly opinion
based and many
times more detailed fact checking is needed. We would like to make more objective clusters of fake, real, and satire news through unsupervised machine learning methods on a variety of news sources. With an expanded
data set and stronger ground truth comes the ability to do more sophisticated classification and more
in-depth natural language feature engineering, all with the hope to
stop the spread of malicious fake news quickly. With more data and
more in-depth features, our arguments could be made much
stronger. Second, we would like to conduct user studies to more directly capture the persuasion mechanisms of fake and real news. For users studies such as this to be valid, careful planning and ethical considerations are needed. Largely, we hope this work helps the academic community
continue to build technology and a refined understanding of malicious
fake news.

\section{Acknowledgments}
{\small Research was sponsored by the Army Research Laboratory and was
accomplished under Cooperative Agreement Number W911NF-09-2-0053 (the
ARL Network Science CTA). The views and conclusions contained in this
document are those of the authors and should not be interpreted as
representing the official policies, either expressed or implied, of
the Army Research Laboratory or the U.S. Government. The
U.S. Government is authorized to reproduce and distribute reprints for
Government purposes notwithstanding any copyright notation here on.}

\bibliographystyle{aaai}
\bibliography{references}

\end{document}